\documentstyle[12pt]{article}
\newcommand{\be}{\begin{equation}}
\newcommand{\ee}{\end{equation}}
\newcommand{\ba}{ \begin{eqnarray} }
\newcommand{\ea}{ \end{eqnarray} }
\newcommand{\bc}{\begin{center}}
\newcommand{\ec}{\end{center}}
\newcommand{\li}{$^{11}$Li\ }
\newcommand{\lib}{$^{10}$Li\ }
\newcommand{\lin}{$^9$Li\ }
\newcommand{\cu}{$^{63}$Cu\ }
\newcommand{\al}{$^{27}$Al\ }
\newcommand{\ca}{$^{12}$C\ }
\newcommand{\pb}{$^{208}$Pb\ }
\newcommand{\s}{2s$_{1/2}$\ }
\newcommand{\p}{1p$_{1/2}$\ }
\newcommand{\sir}{$\sigma_R$\ }
\newcommand{\sinn}{$\sigma_{-2n}$\ }
\newcommand{\sit}{$\tilde{\sigma}_R$\ }
\newcommand{\sid}{$\Delta \sigma_R$\ }

\newcommand{\sinr}{$\sigma^N_R$\ }
\newcommand{\sic}{$\sigma^C_{-2n}$\ }
\newcommand{\sinb}{$\sigma^N_{-2n}$\ }

\newcommand{\plb}[1]{Phys. Lett. {\bf #1}}

\newcommand{\prle}[1]{ Phys. Rev. Lett. {\bf 1}}
\newcommand{\pr}[1]{Phys. Rev. {\bf #1}}
\newcommand{\apb}[1]{Ann. of Phys. (N.Y.) {\bf #1} }
\newcommand{\npb}[1]{Nucl. Phys. {\bf #1}}
\newcommand{\bi}{\bibitem}
\begin{document}

 \begin{flushright} 
IFUP-TH 29/2000
 \end{flushright}

\begin {center}{\Large Reaction and break-up cross sections of \li at 0.8 and 0.28
GeV/u.}\end{center}
 \begin {center}{ \large A. Bonaccorso }\end {center}
\begin {center}{  Institute for Nuclear Theory, Seattle WA, 98195-1550, USA}
\end{center}
\begin {center}{Istituto Nazionale di Fisica Nucleare, Sezione di Pisa, 56100
  Pisa, Italy}
\footnote{Permanent address. Electronic address angela.bonaccorso@pi.infn.it}\end{center}
 \begin {center}{and}\end{center}
 \begin {center} {\large N. Vinh Mau }\end{center} \begin {center}{\normalsize Institut de
Physique Nucl\'eaire, F-91406 , Orsay Cedex, France }\end{center}

\begin{abstract} 
In this paper we calculate reaction and breakup cross sections for the two-
neutron halo nucleus of $^{11}$Li using the optical limit of Glauber theory.
Calculations are presented and compared to experimental data at 0.8 and
0.28 GeV/u on a series of targets. The  $^{11}$Li nucleus 
is described as a three-body system, a core plus two neutrons, with a
phenomenological neutron-core potential and a density dependent
neutron-neutron interaction of zero range. Three different wave functions
are constructed which have different  $(2s_{1/2})^2$ and $(1p_{1/2})^2$ 
two-neutron components but correspond to the same binding energy close to the
experimental value. We show that the  agreement with all the
experimental observables is achieved only if the \li wave function contains
about 30\%
of $(2s_{1/2})^2$ configuration. 
 
\end{abstract}
\section{Introduction}
In two-neutron halo nuclei the reaction and two-neutron 
break-up cross sections are much
larger than in normal nuclei because of their very low two-neutron
separation energy and large radius. These two observables are thus very
useful tools to investigate properties of halo nuclei. Numerous works,
experimental and theoretical, have been devoted to reactions with a \li
projectile. For long time there was some ambiguity about the structure of
\li due to the lack of information about the unbound nucleus of \lib. In
precursor experiments a
p$_{1/2}$ neutron resonance at 0.8 MeV was assumed to be the ground state of
\lib and first theoretical works assumed that the ground state of \li
was mainly formed of
two neutrons in a p$_{1/2}$ resonance. Now it is well accepted that the
ground state is nearly bound with an energy of 0.1-0.2 MeV
\cite{am,bo,yo,kr,zin1,zin2,cha,thoe} and that it is a l=0 
state while the p$_{1/2}$ 
resonance is
an excited state at an energy ranging from 0.35 to 0.6 MeV, implying that the
correlated two halo-neutron wave function in \li will involve both  (2s$_{1/2}$)$^2$
and (1p$_{1/2 })^2$ two-neutron components.
However for a given neutron-neutron interaction, calculations in
Faddeev \cite{tho} or pairing \cite{pvm,beh} model show that the measured
two-neutron separation energy in \li can well be reproduced with different
positions of s$_{1/2}$ and p$_{1/2}$ neutron states therefore with different
(\s)$^2$ and (1p$_{1/2}$)$^2$ components in the wave function. 
Then one needs other information to
discriminate between the different scenarii. Different mixtures in the wave
function correspond
to different radii and comparison of calculated and measured radii gives a
first indication. However the determination of the experimental radius  depends
on the reaction model and on the assumed \li density 
while the calculated radius seems to depend on the structure microscopic 
model. This will be discussed in the present work. Other information comes from
the presence of a low energy dipole mode in \li and the measured B(E1) which
has been shown to favor a wave function with 30-40\% of (2s$_{1/2}$)$^2$
configuration \cite{thz,bvm}. Reaction and break-up cross sections are
expected to be sensitive to both, radius and configuration mixing, and then
to give further constraints on the structure of \li. 

In the present work we study reaction and two-neutron removal cross
sections in the optical Glauber approximation describing the \li nucleus in
a   two-neutron model. The  interaction is assumed to be density dependent  with
zero range. Two of the three parameters of the interaction
are taken from the work of Schuck et al \cite{sch} while the third one,
the strength of the density independent term,  is fitted on the
two-neutron separation energy in
$^{14}$C  which has the same number of neutrons as \li. 
Then varying the energies of the 
s$_{1/2}$ and p$_{1/2}$ states in \lib we look for a two-neutron binding energy
 in \li close to the experimental value. This can be obtained for
different couples of neutron states but leads to different mixtures of
(2s$_{1/2}$)$^2$ and (1p$_{1/2}$)$^2$ configurations in the wave function, thus
to different \li radii. We have chosen three such situations, calculated the
corresponding cross sections for \li +\ca reactions at 0.8
GeV/u and show that we are able to discriminate between these 
different \li wave
functions. Using the wave function which reproduces at best the measured
cross sections for this system we consider other targets and make the
calculation of cross sections at 0.8 and 0.28 GeV/u incident energies. 
Numerous theoretical works
have reported on calculations of cross sections but most of them assumed 
two independent halo-neutrons in a  (1p$_{1/2}$) state. In the present 
paper we shall compare our results mainly with the three most recent works by
Al-Khalili et al. \cite{tha}, Bertsch et al. 
\cite{bhe} and Garrido et al. \cite{gar}   who use correlated wave functions. 

  This paper is organized as follows: in  Section II we introduce the optical 
 limit of Glauber reaction  model and  give the detailed expressions of the
 cross sections; Section III deals with the choice of the numerical 
 parameters entering the calculations; Section IV
presents the results for reaction and breakup cross sections  at the incident
energies of 0.8 and 0.28GeV/u on the targets
\ca,\al,\cu and \pb; finally in  Section V we give our conclusions.

\section{Model of reaction}

In a Glauber eikonal model \cite{rjg} the reaction cross section for nucleus-nucleus
collisions is:
\be
\sigma_R=\int d^2{\bf b}\;(1-P({\bf b}))\label{1}
\ee
where {\bf b} is the impact parameter of the projectile relative to the
target and P({\bf b}) the probability that the projectile passes through the
target without interacting. P({\bf b}) is related to the Glauber phase by:
\be
P({\bf b})=\exp \left (-2Im \chi({\bf b})\right )
\label{2}\ee

In the optical limit \cite{cziz} with a t-matrix approach for the
projectile-target effective interaction, the phase $\chi$ is simply related
to the nucleon-nucleon profile function $\gamma_{NN}$ by:
\be
\chi({\bf b})\simeq\chi_0({\bf b})=i\;\int \int d{\bf r_i}\;d{\bf r_j}\;
\rho_p({\bf r_i})\;\gamma_{NN}(|{\bf {s_i-s_j-b}}|)\;\rho_t({\bf r_j})
\label{3}\ee
$\rho_p$ and $\rho_t$ are respectively the projectile and target densities
and {\bf s} is the projection of the 3-dimensional coordinate {\bf r} on the
plane perpendicular to the z-axis. The profile function is defined in terms
of the two-dimensional Fourier transform of the nucleon-nucleon scattering
amplitude  $f_{NN}(q)$ by:
\be
\gamma_{NN}({\bf s})=\frac{1}{2\pi ik_{NN}}\int exp(-i{\bf q\cdot s})\;f_{NN}({\bf
  q})\;d{\bf q}
\label{4}\ee

while the scattering amplitude can be parametrised as:
\be
f_{NN}({\bf
  q})=\frac{k_{NN}}{4\pi}\;\sigma_{NN}(i+\alpha_{NN}(q))\;exp(-q^2r_0^2/4)\label{5}
\ee
where $\sigma_{NN}$ is the average nucleon-nucleon cross section and $r_0$
has been defined to be the range of the profile function.

For spherically symmetric target and projectile the equations (2) to (5)
lead to a probability P(b):
\be
P(b)=exp(-{\sigma_{NN}\over 2\pi} \int_0^\infty
q\;dq\;\rho_p(q)\;\rho_t(q)\;e^{-q^2r_0^2/4}\;J_0(qb))\label{6}
\ee
with $\rho_{p(t)}(q)$ the 3-dimensional Fourier transform of the
projectile(target) ground state density defined below in eq.(\ref{9}).

Our projectile of \li will be described as a core plus two valence neutrons
in a two-neutron pairing model where the ground state wave function  is
assumed have the form:
\be
\Psi_0=\Phi_0({\bf 1,2})\cdot \Phi_c.\label{7}
\ee
Here $\Phi_c$ is the core wave function and $\Phi_0$ the
correlated two-neutron wave function. The coordinates
{\bf 1,2} include spin as well as spatial ({\bf x}$_i$)
coordinates of the valence neutrons. The coordinates {\bf x}$_i$ are defined relative to the core center of mass. 

In eq.(\ref{6}) the density $\rho_p$ is expressed in terms of coordinates
${\bf r}_i$ relative to the center of mass of the projectile as shown in
Fig.1. By definition the density is:
\be
\rho_p({\bf r})=<\Psi_0({\bf{1,\ldots ,A}})|\sum_{i=1}^A \delta({\bf{r-r_i}})|
    \Psi_0({\bf{1,\ldots ,A}})>\label{8}
\ee 
Using the relations between ${\bf r}_i$ and {\bf x}$_i$ coordinates 
we may write the 3-dimensional Fourier transform of the \li density with respect to ${\bf q}\equiv(q_x,q_y,0)$ as:
\ba 
\rho_p(q) &=&\int\;d{\bf r} \;e^{i{\bf
     q\cdot r}}\;<\Psi_0|\sum_{i=1}^A\;\delta({\bf r-r_i})|\Psi_0> \label{9}\\
       &=&\sum_{i=1}^A\;<\Psi_0 |\exp
       (i{\bf{q}}\cdot ({\bf {x_i}}-\frac{\bf{x_1+x_2}}{A})|\Psi_0>\label{10}
\ea 

Inserting eq.(\ref{7}) in eq.(\ref{9}) we get:
\ba
 \rho_p(q)&=&\tilde{\rho}_c(q)+\tilde{\rho}_h(q) \label{11}\\
\tilde{\rho}_c(q)&=&\rho_c(q)\;\int\;\int \,d{\bf x_1}\;d{\bf
  x_2}\;\rho({\bf{x_1,x_2}})\;exp(-i{\bf{q}}\cdot \frac{{\bf{x_1+x_2}}}{A})\label{12}\\
\tilde{\rho}_h(q)&=& 2 \int\;d{\bf x_1}\;d{\bf x_2}
  \rho({\bf{x_1,x_2}}) exp(-i{\bf{q}}\cdot ({\bf{x_1}}-\frac{\bf{x_1+x_2}}{A}))\label{13}\ea
where $\rho_c(q)$ is  the Fourier transform of the core density
expressed in its own
center of mass system and $\rho({\bf x_1,x_2})$, the two-neutron density
normalized to 1, is defined as:
\be
\rho({\bf x_1,x_2})=\sum_{spins} |\Phi_0({\bf 1,2})|^2\label{14}
\ee

Eq.(\ref {11}) inserted in eq.(\ref{6}) gives:
\be
P(b)=\tilde{P}_c(b)\;P_h(b)\label{15}
\ee
where $\tilde{P}_c $ and $P_h$ are given by eq.(6) with $\rho_p$ 
replaced by $\tilde{\rho}_c$ and $\tilde{\rho}_h$ respectively. 

Inserting eq.(\ref{15}) in eq.(\ref{1}) we may write \sir  as:
\ba
\sigma_R(^{11}Li)&=&2\pi\int\;b\;db\;(1-\tilde{P}_c\;P_h) \label{16}\\
        &=&2\pi\int\;b\;db\;((1-\tilde{P}_c)+(\tilde{P}_c\;(1-P_h)) \label{17}\\
        &=&\tilde{\sigma}_R(^9Li)+\sigma_{-2n}\label{18}
\ea
where $\tilde{\sigma}_R$(\lin) is the core contribution to \sir(\li) but is
different from the \lin-core reaction cross section due to the core recoil.
 Indeed $\tilde{\rho}_c$ of eq.(\ref{12}) can be written in a more explicit way by
 introducing ${\bf r}_{12}$, the distance between the two halo-neutrons and
 ${\bf r}_{cm}$, the distance between the \li and \lin centers of mass
 related to ${\bf x}_1$ and ${\bf x}_2$ by:
\be
{\bf r}_{cm}=\frac{{\bf x}_1+{\bf x}_2}{A}\;\;\;\;{\bf r}_{12}=
{\bf x}_1-{\bf x}_2\label{19}
\ee
To simplify the  equations we ignore spin variables and
write, following eq.(\ref{14}):
\be
\rho({\bf x}_1,{\bf x}_2)=|\Phi_0({\bf x}_1,{\bf x}_2)|^2\equiv |\Phi({\bf
  r}_{cm},{\bf r}_{12})|^2\label{20}
\ee
Then eq.(\ref{12}) can be transformed into:
\be
\tilde{\rho}_c(q)=\rho_c(q)\;\rho_{cm}(q)\label{21}
\ee
where $\rho_{cm}(q)$ is the Fourier transform of the density distribution of
the core center of mass motion relative to the \li center of mass  given by:
\be
\rho_{cm}(r_{cm})= \left (\frac{A}{2} \right )^2\;\int d{\bf r}_{12}|\Phi({\bf r}_{cm},
{\bf r}_{12}|^2\label{22}
\ee
Eq.(\ref{17}) shows that
$\tilde{P}_c$ can
be interpreted as the probability that the core, inside the projectile, has
no interaction with the target while $\tilde{P}_h$ concerns  the two
halo-neutrons.

The second term of eq. (\ref{18}), which we are calling $\sigma_{2n}$ represents
dominantly valence particle effects. It is  the contribution of the two neutrons  to the
reaction cross section  and gives the  two-halo-neutron removal cross section, including
break-up, neutron transfer to the target and inelastic processes.

Note that if $\frac{{\bf{x_1+x_2}}}{A}$ is a small quantity, namely if the
core and projectile centers of mass are close, the exponential $exp(-i{\bf 
  q}\cdot \frac{{\bf x_1}+{\bf x_2}}{A})$  in eqs.(\ref{12})  and (\ref{13}) can
be replaced by  1 and we can write:
\ba
\tilde{\rho}_c(q)&\simeq& \rho(q)\\
\tilde{\rho}_h(q)&\simeq &2\;
\int\;d{\bf x_1}\;d{\bf x_2}\;\rho({\bf{x_1,x_2}})\;exp(-i{\bf{q\cdot x_1}})\\
   &=&2\;\rho_n(q)
\ea
where $\rho_n(q)$ is the Fourier transform of the one-neutron average
density given by:
\be
\rho_n({\bf x})=\int d{\bf x'}\;\;\rho({\bf{x,x'}})\label{26}
\ee

In this limit which is valid for normal (or heavy) nuclei we get:
\be
\sigma_R=\sigma_R(^9Li)+\;\sigma_R(2n)\label{27}
\ee
then the reaction cross section is the sum of two independent cross sections
for the core and the two valence-neutrons. This however does not hold for
\li where the two halo-neutrons are far away from the core.

Before ending this section we discuss the validity of the reaction model presented above.
Equation (\ref{3}) is based on the leading order term in  the cumulant expansion
of the multiple scattering series \cite{fv} for the eikonal operator:
 $\int d^3r \rho_{2n}exp(-\sigma_{NN}\int\rho_T dz^{\prime})\approx
exp(-\sigma_{NN}\int\rho_{2n}\rho_Td^3r dz^{\prime})$.
According to Yabana et al.\cite{ia} this 
    is not very well justified in the intermediate energy region and for extended halo
nucleon wave functions. However its   use can be justified in the cases studied in this
paper  in view of the fact that we work
 in the hight energy regime and that the separation
between valence and core particles cannot be done exacly working in the
center-of-mass of a halo projectile nucleus as we do in the present work. Furthermore, as we shall see in the 
following our numerical results are quite close to  those of Bertsch et al. \cite{bhe} who do not use the cumulant expansion but
on the other hand  use a purely imaginary nucleon-nucleon scattering amplitude as we do here. A second drawback of the 
 optical limit  giving   eq.(3) is the fact that only the imaginary part of
the nucleon-nucleon scattering amplitude enters the calculation.  The eikonal phase shift eq.(3)
 is thus equivalent to a phase shift due only to an imaginary nucleus-nucleus
 optical potential. At high energy where
$\alpha_{NN}$ of eq.(5) is small this is a good approximation. 
However at 0.28 GeV per
nucleon $\alpha_{NN}$ is small but might not be negligible. Ray \cite{ray} 
 gives $\alpha_{PN}=0.16$ at 325MeV. 
Thus Bertsch et al.\cite{bhe} neglected it. On the other hand  Garrido et al. \cite{gar}  
used a phenomenological optical potential with
both real and imaginary parts for the neutron target scattering. 
 At this point an important remark is in order. As it has been discussed in 
\cite{bob,bc} and references therein, the neutron target interaction optical potential 
varies  from light to heavy targets and
increasing the incident energy reflecting the change in the reaction mechanism.
In particular the relative amount of real and imaginary parts can be rather different
 from one target to the other leading to different amount of neutron elastic and inelastic
scattering. This in turn is reflected in rather different amounts of the so called
diffraction and absorptive breakup from a halo projectile.
 Around 300MeV experimental data \cite{exp} show
that the total n+$^{12}$C cross section is largely dominated by the reaction cross section. 
The same is not true for a Pb target,
for example, where, at the same energy, reaction and elastic free neutron cross sections
 are of comparable magnitude. We conclude
that our calculated cross sections at the lower energy might underestimate the measured 
cross sections because the so called
diffraction component of the breakup cross section, corresponding to the neutron elastic 
rescattering on the target \cite{bob} is
calculated only with the imaginary part of the nucleon-nucleon amplitude.

\section{Inputs of the calculations}

We apply the reaction model of the previous section to \li on \ca,\al,\cu and
$^{208}$Pb for incident energies of 0.8 and 0.28 GeV/u. To perform the
calculations we need  the nucleon-nucleon parameters, average cross section
and range of the profile function, target densities and  the \li wave function.

\subsection{Nucleon-nucleon parameters}

The nucleon-nucleon cross section $\sigma_{NN}$ of eq.(\ref{5})
 averaged over the neutron-neutron $\sigma_{nn}$, proton-proton $\sigma_{pp}$
and neutron-proton $\sigma_{np}$ pairs is taken as defined
by Charagi and Gupta \cite{ch} and calculated for each target from their
parametrisation of $\sigma_{pp}(=\sigma_{nn})$ and $\sigma_{np}$ 
at 0.8 and 0.28 GeV.
As they  depend only very slightly on the targets, we use a unique
value $\sigma_{NN}$= 4.1 and 3.1 fm$^2$ for incident energies of 0.8 and 0.28
GeV/u respectively. 

The average range of the nucleon-nucleon profile function, eq.(\ref{4}), is
extrapolated from the tabulation given by Ray \cite{ray} and found to be 0.64
and 1.41 fm at 0.8 and 0.28 GeV respectively. The value of 0.64 fm is
close to the values used at high energy by Charagi and Gupta
\cite{ch} and Cziz and Maximon
\cite{cziz}. In most of the calculations for a \li projectile a zero range is
assumed but we shall see that this choice has some effect on 
the determination of the $^{9}$Li radius.

\subsection{Target densities}

For a \ca-target we use a Gaussian density with a range fitted to reproduce
the radius of 2.32 fm. We have checked that an harmonic oscillator density
gives the same cross sections. 
For the heaviest targets we take a Fermi density with
parameters determined from electron scattering \cite{sick} and neglect
differences between neutrons and protons. For \pb we have also used the
theoretical density of Brack et al.\cite{bra} with different parameters for neutron and
proton densities but the difference between the results on cross sections
are very small compared to the uncertainties due the reaction model and to 
the structure model used to describe \li. 
 \subsection{\li wave function}

We construct several wave functions following ref.\cite{pvm}. We replace continuum
states by
discrete states calculated in a radial box of radius 20 fm and take all
neutron states up to an energy of 8 MeV. Instead of a Woods-Saxon
neutron-core potential with a strength fitted separately 
on \p and \s neutron energies,
as usually done, we take an usual Woods-Saxon potential with fixed
parameters and correct it for the two low energy \s and \p resonances in $^{10}Li$ 
by a surface term due to neutron-core
vibrations coupling and fitted to each resonance \cite{pvm,nvm}. 
These two choices of neutron-core potential
are not equivalent since the surface term in the second choice modifies the
radius of the potential without changing its strength. Our modified average
neutron-core potential  has been shown to give simultaneous good description
of the two mirror nuclei  $^{11}$Be and $^{11}$N \cite{svm}.

The  neutron-neutron effective interaction is chosen of simple form:
\be
V_{nn}(1,2)=-\left (V_0-V_{\rho}\;\;\left (\frac{\rho_c(r_1)}{\rho_0}\right )^p\right ) \delta({\bf {r_1-r_2}})
\label{28}\ee
where $\rho_c(r)$ and $\rho_0$ are the core and nuclear matter densities
respectively. In our first papers on \li structure we have taken  
 p=1.2 and a constraint $V_0\simeq V_{\rho}$. The strengths
$V_0$ and $V_{\rho}$ were then fitted to $^{14}$C and $^{12}$Be two-neutron
separation energies. However the determination of the three parameters of
eq.(\ref{28})      is not unique and to restrict the number of parameters we have
taken the two parameters p and $\alpha=V_{\rho}/V_0$ determined by Garrido
et al.
\cite{sch} in order to reproduce the nuclear 
matter gap calculated with the Gogny
effective pairing interaction. To get agreement in all domain of $k_F$ they
have to assume p=0.47 and $\alpha$=0.45, then very different parameters
compared to what is usually employed \cite{pvm,beh,bf}. With a similar
adjustment, close parameters have been found by Bertsch and Esbensen
\cite{be}. 
Taking these two parameters we
have determined the third one, $V_0$, in order to reproduce closely the
two-neutron separation energy in $^{14}$C. This gives $V_0$ =890 MeV.fm$^{3}$
 ($V_{\rho}$=440 MeV.fm$^{3}$). 

To construct different wave functions we keep the effective pairing
interaction fixed and vary the \p and \s neutron energies simultaneously
in order to get the same two-neutron separation energy S(2n)=0.36 MeV. This
value corresponds to the highest value compatible with measurements but we
have checked that fixing S(2n)=0.32 MeV as in ref.\cite{tho} or 0.295 MeV as in
ref.\cite{bhe} would not change our conclusions. 
This way of deriving different wave functions is similar to Thompson and Zukhov \cite{tho}
but different from Esbensen et al. \cite{bhe} who fix the \p energy and vary both 
the \s energy and the strength of the density dependent term 
of the effective interaction.
We present  in Table I
three such typical wave functions, called F$_1$, F$_2$ and F$_3$,
with the corresponding \p and \s neutron
energies, the rms-radius of \li assuming the \lin  radius to be 2.32 fm,
the percentages of (\s)$^2$ and (\p)$^2$ configurations and the mean
distance  between the core and two halo neutrons centers of mass.
 This last quantity will tell us about the validity of the optical Glauber
 approximation. We see that for \s and \p energies close to the most
 recent experimental neutron energies the F$_2$ wave function has a
large(27\%) component of (\s)$^2$ two-neutron state with a \li radius
compatible with the value determined from reaction cross sections by
Tanihata et al. \cite{tan} but smaller than the value given in a more recent analysis of
Al-Khalili et al.\cite{tha}. This question will be discussed in more detail in
section IV. 

\section{Results and discussion}

From the equations of section II and the inputs of the previous section we
calculate the reaction and two-neutron removal cross sections for a \li
projectile on different targets at two incident energies of 0.8 and 0.28
GeV/u. We first study \li on \ca at 0.8 GeV/u. Indeed for a \ca target the
Coulomb contribution to cross sections is negligible and we can compare
directly the calculated cross sections of eqs.(\ref{18}) with measurements. On the
other hand at such a high energy the validity of the optical Glauber
approximation is better. At 0.8 GeV/u it has been evaluated in a simple model
by Tostevin et al. \cite{alt} and shown to overestimate the reaction
cross section by 1 to 5  \% for the domain of distances between the core and
two-neutron centers of mass implied by our \li wave functions of Table I. It
is well beyond what can be expected from nuclear structure models. We calculate
$\sigma_{R}$ and $\sigma_{-2n}$ for our three wave functions and keep the
one which reproduces at best the measurements.
Then with this 'best' wave function
we report on our results for the other targets at the two incident energies.
At 0.28 GeV/u our results represent only a part  of the breakup cross section
since, as discussed in
section II, the diffractive contribution to the cross sections is calculated 
only via an imaginary potential.

Before going to \li we look at the reaction \lin+\ca at 0.8 GeV/u and
investigate the sensitivity of the extracted rms radius of \lin 
to the range of the
nucleon-nucleon profile function.

\subsection{\lin+\ca at 0.8 GeV/u and \lin radius}

 We take the \lin core density  as a Gaussian  \footnote{ We have checked that
 a harmonic oscillator density leads to the same results} and fit the radius
to reproduce the measured reaction cross section
$\sigma_R$=796 $\pm$ 6 mb \cite{tan}. With r$_0$, the range of the n-n profile
function, equal to zero one finds $<r^2>^{1/2}$= 2.32 fm for \lin as already
given by Tanihata et al.\cite{tan}. Taking r$_0$= 0.64 fm we have to assume a radius of
2.18 fm to recover the experimental cross section of 796 mb. This new value
is significantly smaller  and in fact in better
agreement with the variation as A$^{1/3}$ compared to the \ca radius for
example. It shows that what we call the experimental radius 
is in fact model dependent.

In the rest of the paper we use for definiteness r$_0$=0.64 fm and a radius of
2.18 fm for \lin.

\subsection{\li+\ca at 0.8 GeV/u}

Our cross sections are summarized in Table II for the wave functions F$_1$,
F$_2$ and F$_3$ of Table I. We give the three quantities,
\sit(\lin),$\sigma_R$(\li) and $\sigma_{-2n}$(\li). We see that
\sit(\lin), the core contribution to the \li reaction cross section of eq.(\ref{18}),
is always larger that the reaction cross section for a \lin projectile implying
that \sinn is always smaller than \sid, the difference between \li and \lin
reaction cross sections, in agreement with experimental observation when
available. Comparing our results with the measured values \cite{tan,kob}
for the three
quantities \sir, \sid and \sinn simultaneously we see that an overall
good agreement between theory and experiment is achieved for the F$_2$ wave
function only. This wave function has 27\% of (\s)$^2$ and 52\%  of (\p)$^2$ 
configurations respectively. The F$_1$  wave function which has
 39\% of (\s)$^2$ and the F$_3$ wave function  which has instead 80\% of

\noindent (\p)$^2$
configuration  give too large and too small cross sections.

 At this point it is instructive to compare our wave function F$_2$ to wave
 functions derived in other works and shown to give the best results for
 reaction or two-neutron removal cross sections. We again point out that
 our approximation of working with an S matrix for the whole \li instead of
 a product S$_c$S$_{n_1}$S$_{n_2}$ implies, following ref.\cite{tha},  for F$_2$ which
 corresponds to $<r^2_{c,2n}>^{1/2}$=4.5 fm an overestimate of few \%  which
 leaves our results within the experimental error bars. In a Faddeev model,
 Thompson and Zukhov \cite{tho} get two wave functions, P$_2$ and P$_3$,
 which lead to
 agreement with respectively the minimum and maximum of the measured
 reaction cross sections for
 \li+\ca at 0.8 GeV/u \cite{tha,alt}. For breakup at the same energy 
Bertsch et al.\cite{bhe}
 et al obtain also for their two-neutron wave function S$_{23}$  of ref.[11] 
 slightly too
 small   cross section; considering other observables they conclude
 that the best mixture of (\s)$^2$ state should be around 30-40\%.

 The properties of all these
 wave functions are summarized in Table III. Following the above papers 
 we give the neutron-core s-state scattering length rather than
 the \s energy. We
 note that the percentage of (\s)$^2$ component is similar in the four
 wave functions. The wave functions P$_2$, P$_3$ and F$_2$ are constructed in different
 structure models but with similar neutron energies, a low \s state and
 a \p resonance at 0.25-0.35 MeV lower than given by measurement. 
 The wave function 
 S$_{23}$ however corresponds to higher \s and \p states.
 We see from the table that for similar
  neutron energies and mixtures the radius calculated in Faddeev model is
  larger than the radii given by pairing model. Such a discrepancy
  could have several
 origins. Concerning pairing model, one can invoke the discretization of
 continuum used in ref. \cite{pvm,beh}  and in this work. Indeed the discrete, positive
 energy states are calculated in a box of finite radius and the final \li
 radius
 could depend on the radius of the box. This approximation is currently used
 in normal heavier nuclei where continuum has little effect on ground state
 properties. In \li where all neutron  states are unbound this
 approximation may be suspected. An indication of the effect of the
 radial cut-off could be found by comparing the radii of F$_2$ and S$_{23}$
 since they were calculated in a box of 20 and 40 fm respectively. Indeed
 F$_2$ has a  smaller radius  but the radius depends also on the single neutron
 energies which are  different in the two works
 and perhaps on the energy cut-off
 which is higher for S$_{23}$. An example of a possible disease
 coming from continuum discretization can be found in $^6$He where, using
 experimental neutron resonance energies in $^5$He and the same pairing
 force as in \li, a pairing model with continuum discretization 
 gives  too large binding due to strong couplings of the resonances with
 higher discretized continuum states. A treatment of
 continuum avoiding discretization seems to give weaker couplings 
 with however a larger
 radius \cite{mpv}. These two points have to be confirmed but suggest a
 sensitivity of the radius of two-neutron halo nuclei to the nuclear model
 and, inside the model, to the treatment of the nonresonant continuum
 neutron states. 

As a conclusion of this subsection our analysis shows a great sensitivity of
reaction and two-neutron removal cross sections to the percentage of $(2s_{1/2})^2$
two-neutron configuration and agrees with other analysis in  predicting a
percentage of about 30\% . It is difficult to make a precise prediction
on \li  radius which is strongly model dependent. Besides  the uncertainty
discussed above, there is  an other uncertainty coming from the \lin
radius. For example  a radius of 3.4 fm calculated assuming a radius of
2.32 fm for \lin becomes 3.3 fm if we adopt the value of 2.18 fm deduced above. 
Note also that the removal cross section is more sensitive to the details of the
\li wave function and  is a more stringent constraint than the reaction cross
section. 

\subsection{\li+\al, \cu and \pb at 0.8 GeV/u} 

We present now our cross sections for different targets using the results of the
previous subsection that the best overall agreement is obtained for our wave
function F$_2$ which is used from now on in all our calculations.
The cross sections are presented in Table IV (upper part) and
compared with available measurements \cite{kob}.
In the table the indices N and C mean nuclear
and Coulomb cross sections respectively, the sum of the two being compared 
 with  measurements. For the heavier targets 
Coulomb contributions cannot be neglected. We have calculated \sic
following our previous paper on low energy dipole mode 
and Coulomb break-up \cite{bvm}
which assumes that the two-halo neutrons are ejected simultaneously. The low
energy dipole state and corresponding B(E1) are calculated with the same
neutron energies and effective n-n interaction than used to 
construct the ground state wave function F$_2$
. We see that \sinn=\sinb+\sic is in reasonable agreement with
measurements. For \cu and  \pb targets the total cross section is close to
measurement even though \sic and \sinb differ from the values
extracted by Kobayashi et al \cite{kob} using general properties of the nuclear contribution.
For \al our
\sinr is already close to the largest value indicated by experiment,
suggesting that the Coulomb contribution should be very small, smaller than 
the evaluation given in ref.[30].   This is true  for \sic which is less 
than 5\% of \sinb. 
It is also the case for  \lin+ \al  where our calculated \sir of 1.22 b 
is only slightly larger than the measured one of 1.135$\pm$7 b.

For all targets including \ca, our cross sections are close to the higher
experimental values or even slightly larger but one has to remember that
using the optical limit of Glauber theory leads to an overestimate of
few \%. On the other hand we have to keep in mind that 
nuclear models introduce also some uncertainty
on the final results. 

\subsection{\li+\ca, \al and \pb at 0.28 GeV/u}
At this relatively low incident energy
our optical Glauber approximation which involves only the imaginary part of the
nucleon-nucleon scattering amplitude  does not seem to be  able to describe the whole
two-neutron contribution to the reaction cross section.
However our calculated \sinb should
describe well what is called   the absorption or inelastic or stripping
contribution ($\sigma^S_{-2n}$). We give in Table IV (bottom) our results and
the measured cross sections when available \cite{zin1,zin2,hum}. 
Blank et al \cite{bla} have measured \sir($^9$C+\ca) at 280 MeV/u to be
812$\pm $34 mb. If we assume that $^9$C and \lin have similar density our
value of 842 mb agrees well with this result. 

Our two-neutron removal cross sections are always smaller than the measured
ones as expected.
 Bertsch et al.\cite{bhe}  obtain also for their two-neutron wave function S$_{23}$  of ref.[11] a too small two-neutron removal  cross section. 
However our value \sinn(\li+\ca)=0.2 b is in good agreement with the stripping
contribution 
extracted from the measured \sinn by Zinser et al. \cite{zin2}and with the value obtained by Garrido et al.\cite{gar} with a Faddeev \li wave function built from low energy \s and \p neutron states in \lib.  

It is worth noticing, as it has already been remarked in \cite{bhe}, that
the experimental two neutron removal cross section at 280  MeV/u is larger
than the cross section measured at 800 MeV/u.  One-step breakup  models of the Glauber type  discussed here\cite{bhe,og}
 and based
on the  nucleon-nucleon cross section or  based on the use of a 
phenomenological  nucleon-target optical potential
\cite{gar,ia,bob} would predict an opposite trend.

Finally we notice that Zinser et al \cite{zin2},
with an invariant-mass analysis of \lin + n system after break-up of \li on
the \ca target, 
have extracted energies of s and p states in \lib
and the corresponding relative intensity of (\s)$^2$  and (\p)$^2$
components in \li. Our wave function F$_2$ qualitatively agrees with their
results even though our \p resonance energy is lower. In their analysis they
see a higher broad peak at about 1.6 MeV  assigned as a p-resonance. In our
discrete basis simulating continuum we have a 2\p state at 1.43 MeV which
gives rise to (1p$_{1/2}$,2p$_{1/2}$)  and  (2p$_{1/2}$)$^2$ components in \li with 
 probabilities of 10 and 1.5\% respectively and
which could, perhaps, be identified with this experimental peak. 
However we have not checked that this discrete state corresponds 
to a resonance.

\section{Conclusion}

In this paper reaction and two-neutron removal cross sections have been calculated for a
\li projectile at two
incident energies of 0.28 and 0.8 GeV per nucleon and several targets, using
 the optical approach to Glauber model where the $^{9}$Li-core recoil
has been taken into account.
We have worked with a finite range nucleon-nucleon profile function and
observed that the ``measured'' \lin radius fitted to reproduce the \lin+\ca
reaction cross section at 0.8 GeV/u is quite sensitive to this parameter. With
a range $r_0= $0.64 fm taken from nucleon-nucleon systematic we get
$<r^2>^{1/2}$=2.18 fm instead of 2.32 fm, the commonly accepted value
obtained with $r_0$ =0. This new value is in better agreement with a variation
as A$^{1/3}$ when compared for example to the \ca radius of 2.32 fm determined
from electron scattering.

The \li wave function has been calculated in a two-neutron pairing model
with a zero range and density dependent pairing interaction suggested by a
fit of the nuclear matter gap calculated with a Gogny finite range effective
interaction. We have found that for \ca,\al,\cu and \pb targets
and an incident energy
of 0.8 GeV/u,  reaction and two-neutron removal cross sections are
simultaneously well reproduced for a \li wave function having about 30\% of
(2s)$^2$ configuration. At an incident energy of 280 MeV/u however,  where
what is called diffraction contribution  has been measured to be 30\% of the total removal
cross section, we get too low cross sections because we use  the optical
approach to Glauber model which involves only the absorptive part of the nucleus-nucleus
potential. However our results compare well with the calculated or measured
absorption contribution.

Our finding of about 30\% of (2s)$^2$ component in the \li wave function
agrees with the conclusion of other works on \li+\ca reaction cross section
at 0.8 GeV/u or removal cross section at 0.28 GeV/u. However our theoretical
radius is smaller than that deduced by Al-Khalili et al using three-body
Faddeev model but close to the ``measured'' radius deduced by Tanihata from
reaction cross section assuming a Gaussian or harmonic oscillator \li
density. This reveals a strong dependence of the radius on the structure
model which comes, one may guess, from the large distance part of the wave
function. In our pairing model continuum neutron states are approximated by
discrete states calculated in a radial box, thus with a radial cut-off of
the wave function. If the wave function has large enough components on
two-neutron configurations involving continuum non resonant neutron states,
the rms radius will be possibly underestimated in our
model as well as in an empirical analysis of data using Gaussian or harmonic
oscillator density. In the case of our 'best' wave function F$_2$ such
configurations contribute to about 12\%, thus may  introduce indeed a
sensitivity of the radius to the radial cut-off.      

{\bf Acknowledgements}

We wish to thank George Bertsch for discussions and a critical reading of the manuscript.

\newpage

\begin{table}
\begin{center}
\caption{Binding energies, r.m.s. radii and weights of the (2s)$^2$ and
  (1p$_{1/2})^2$ components in the wave function obtained
  with different single neutron energies.The r.m.s. radius of $^{9}$Li is taken
  as 2.32 fm. }
\begin{tabular}{|c|ccc|c|cc|}\hline
wave &$\epsilon$(1p$_{1/2}$)&$\epsilon$(2s)&S(2n)&
$<r^2>^{1/2}$&(2s)$^2$&(1p$_{1/2})^2$\\
funct.&(MeV)&(MeV)&(MeV)&(fm)&\%& \%\\ \hline
F$_1$&0.41&0.14&0.36&3.30&49&36\\
F$_2$&0.35&1.20&0.36&3.11&27&54\\
F$_3$&0.28&0.56&0.36&2.84&1&80\\ \hline
\end{tabular}
\end{center}
\end{table} 

\begin{table}
\begin{center}
\caption{Reaction and two-neutron removal cross sections defined in the text 
  for $^{11}$Li+$^{12}$C at 0.8 GeV/n calculated with the three wave functions
  of Table I. The experimental values are taken from references 
\protect{\cite{tan}} and
 \protect{ \cite{kob}}. The cross sections are expressed in barns.}
\begin{tabular}{|c|ccc|cc|cc|}\hline
wave f.&
$\tilde{\sigma}_R$($^9$Li)&$\sigma_R$($^{11}$Li)&
$\sigma_R^{exp}$($^{11}$Li)&$\Delta\sigma_R$&
$\Delta\sigma_R^{exp}$&$\sigma_{-2n}$&
$\sigma_{-2n}^{exp}$\\ \hline
F$_1$&0.839&1.127&1.060$\pm$
  0.010&0.331&0.264$\pm$0.016&0.288&0.220$\pm$ 0.010\\
F$_2$&0.832&1.072& &0.276& &0.241&\\
F$_3$& 0.818&1.00& & 0.204& &0.183&\\ \hline
\end{tabular}
\end{center}
\end{table} 
\begin{table}
\begin{center}
\caption{s-wave scattering length, 1p$_{1/2}$ single neutron energy, r.m.s.
  radius of $^{11}$Li and weights of (s)$^2$ and (p$_{1/2}$)$^2$
components for wave functions P$_2$, P$_3$ of ref.\protect{\cite{tho}},
 S$_{23}$ of
 ref.[11]  and F$_2$ of this work. The * means that 54\%  is  the
 percentage of (\p)$^2$ state only, the sum over all (p$_{1/2}$)$^2$ states
 is 65\%.}
\begin{tabular}{|c|ccccc|} \hline
& a$_0$&$\epsilon$(1p$_{1/2}$)&$<r^2>^{1/2}$&(2s)$^2$ & (1p$_{1/2}$)$^2$\\
&(fm)&(MeV)&(fm)&(\%)& (\%) \\ \hline
P$_2$&-18&0.25&3.39&31&64\\
P$_3$&-27&0.3&3.64&45&51\\
S$_{23}$&-5.6&0.54&3.22&23.1&61\\
F$_2$&-11.7&0.35&3.11&27&54$^{*}$\\ \hline
\end{tabular}
\end{center}
\end{table}
\setlength{\oddsidemargin}{-1cm}
\begin{table}
\begin{center}\caption{Cross sections defined in the text at two incident energies and for
  several targets  expressed in barns. Experimental values are
  taken from references \protect{\cite{zin2}} for $^{a)}$ and 
\protect{\cite{hum}} for $^{b)}$.}
\begin{tabular}{|c|c|cccc|cccc|} \hline
energy&target&$\sigma_R^N$($^9$Li)&$\tilde{\sigma}_R^N$($^{9}$Li)&
$\sigma_R^N$($^{11}$Li)&
$\sigma_R^{exp}$($^{11}$Li)&$\sigma^N_{-2n}$&$\sigma^C_{-2n}$&
$\sigma_{-2n}$&$\sigma^{exp}_{-2n}$\\ \hline
0.8
GeV/n&$^{12}C$&0.796&0.832&1.07&1.06$\pm$0.01&0.240&-&0.240&0.220$\pm$0.010\\
 & $^{27}$Al&1.22&1.26&1.62&1.56$\pm$0.04&0.361&0.017&0.378& \\
 & $^{63}$Cu&1.83&1.88&2.41&2.55$\pm$0.22&0.527&0.080&0.607&0.52$\pm$0.04\\
 & $^{208}$Pb&3.11&3.19&4.03&5.38$\pm$0.64& 0.840&0.580&1.42&1.31$\pm$0.1\\
 \hline
0.28 GeV/n&$^{12}$C&0.842&0.870&1.07& &0.200&-&0.200&0.28$\pm$0.03$^{a)}$\\
 & $^{27}$Al&1.27&1.31&1.61& &0.300&0.025&0.325&0.47$\pm$0.08$^{b)}$\\
 & $^{208}$Pb&3.26&3.33&4.0& & 0.679&0.823&1.502&2$\pm$0.5$^{a)}$\\ \hline
\end{tabular}
\end{center}
\end{table}

\end{document}